\RequirePackage[2020-02-02]{latexrelease}
\documentclass[12pt]{revtex4}

\usepackage{multirow}
\usepackage{amsmath,amsfonts}
\usepackage{amssymb,latexsym}
\usepackage{color}
\usepackage{graphicx}
\usepackage{mathrsfs}

\newtheorem{prop}{Proposition}

\newcommand{\beprop}{\begin{prop}}
\newcommand{\enprop}{\end{prop}}
\newcommand{\bprf}{\begin{proof}} 
\newcommand{\eprf}{\end{proof}\qed}

\definecolor{hervecolor}{rgb}{0.8,0,0.7}

\newcommand{\ket}[1]{|\kern.3ex#1\kern.3ex\rangle}
\newcommand{\bra}[1]{\langle\kern.3ex #1 \kern.3ex|}
\newcommand{\scalar}[2]{\langle\kern.3ex #1 \kern.3ex|\kern.3ex#2\kern.3ex\rangle}

\newcommand{\ud}{\mathrm{d}}

\newcommand{\uE}{\mathrm{E}}

\begin{document}

\title{Thermodynamical Relations in Function Space}
\author{T.\ Koide}
\email{tomoikoide@gmail.com,koide@if.ufrj.br}
\affiliation{Instituto de F\'{\i}sica, Universidade Federal do Rio de Janeiro, C.P. 68528,
21941-972, Rio de Janeiro, RJ, Brazil}
\author{T. Kodama}
\email{kodama.takeshi@gmail.com,koide@if.ufrj.br}
\affiliation{Instituto de F\'{\i}sica, Universidade Federal do Rio de Janeiro, C.P. 68528,
21941-972, Rio de Janeiro, RJ, Brazil}
\affiliation{Instituto de F\'{\i}sica, Universidade Federal Fluminense, 24210-346,
Niter\'{o}i, RJ, Brazil}
\begin{abstract}
We formulate thermodynamical relations based on the field degrees of freedom 
by introducing a work induced by the volume change in the function space, in addition to the usual work associated with the spatial volume change.
The first and second laws of thermodynamics are defined for such a work inherent to the field theory 
by generalizing stochastic energetics (stochastic thermodynamics) to the classical real scalar field in contact with a heat bath.
We further discuss the local equilibrium ansatz in the function space and show 
that it is possible to introduce a model of functional ideal hydrodynamics, which is consistent with the derived thermodynamical laws. 
\end{abstract}

\maketitle

Hydrodynamics is one of effective theories described in terms of collective variables 
and has been applied to various fields ranging from particle physics to cosmology \cite{hydro_review1,hydro_review2}. 
Several assumptions are utilized in the derivation but we here focus on the local equilibrium ansatz, 
which means that a fluid can be represented by the ensemble of virtual quasi-particles called fluid elements and 
the internal states of these infinitesimal elements are approximately given by thermal equilibrium states. 
This assumption plays an important role in the derivations both of non-relativistic and relativistic hydrodynamics \cite{hugo,degroot}.

At the same time, fundamental degrees of freedom in microscopic systems are known to be described not by particles but by fields. 
Therefore, it is reasonable to expect that hydrodynamics for microscopic systems should be formulated based on the local equilibrium ansatz applied to fields. 
Such a generalization has the advantage that quantum-field theoretical fluctuation can be incorporated to hydrodynamical evolutions \cite{fhydro}. 
For example, in the physics of relativistic heavy-ion collisions, although the collective behaviors of hot matters produced by the collisions are 
consistent with the results of relativistic hydrodynamical models, it is assumed that the quantum effect does not modify the structure of hydrodynamical equations \cite{hydro_review1,hydro_review2}. 
Such a modification is partially taken into account in the study of the trapped cold atoms where dynamics of the Bose-Einstein condensate is well 
described by a hydrodynamical equation obtained by the Madelung transform of the effective equation, the Gross-Pitaevskii equation \cite{vortex_BEC}. 
Higher-order quantum many-body effects which are not considered in the above examples can be studied by applying functional hydrodynamics 
because the exact behavior of quantum field theory is known to be mapped into the form analogous to functional ideal hydrodynamics \cite{fhydro,svm-field,jackiw}.
In addition, functional hydrodynamics may be useful to investigate the collective (hydrodynamical) behaviors of photons where the applicability of the local equilibrium ansatz based on the particle picture is not obvious \cite{photon-hydro}.

To push forward this idea, we have to study thermodynamical relations associated with thermodynamical manipulations in the function space.
For example, we should consider the work associated with the volume change in the function space instead of the usual work characterized by the spatial volume change. 
It is not obvious at all whether we can find the thermodynamical laws to such an abstract manipulation in the function space. 
The framework of thermodynamics, however, has been applied even to abstract concepts like information: 
Maxwell's daemon, Szilard's engine, Landauer's principle and so on. See, for example, Ref.\ \cite{information} and references therein.
Our attempt is an extension of this line of study.

To find the laws analogous to the first and second laws of thermodynamics, stochastic energetics (stochastic thermodynamics) 
is applied \cite{sekimoto,broeck-review,sei-rev,kkRBM,koideQSE,koidereci,pal}.
This approach has been applied exclusively to particle systems, but we now consider the application to a field theoretical system.
As an example, we consider a 1+1 dimensional classical real scalar field in contact with a heat bath of temperature $T$.
In the following discussion, we use the natural unit, $\hbar = c = 1$.
The Hamiltonian is defined by 
\begin{eqnarray}
H (\{\Pi_t,\phi_t\},a_t,b_t) = \int dx\, \left[  \frac{1}{2} \Pi^2 (x,t) + \frac{1}{2} (\partial_x \phi (x,t))^2 + V(x,a_t) \phi^2(x,t) + U (\phi(x,t), b_{t} )  \right] \, , \nonumber
\end{eqnarray}
where $\Pi (x,t)$ is the conjugate field of $\phi (x,t)$ and 
$V(x,a_t)$ is an external confinement potential. 
The width of the potential, which characterizes the spatial volume of this system, 
is controlled by a time-dependent parameter $a_t$.
In a similar fashion, the field interaction $U (\phi(x,t), b_{t} )$ 
contains a term which controles the fluctuation of the scalar field through another external parameter $b_{t}$.

In stochastic energetics, the interaction with the heat bath is taken into account through the thermal relaxation force and the thermal noise.
The evolution equations of $\phi(x,t)$ and $\Pi(x,t)$ are then given by the stochastic differential equations,
\begin{eqnarray}
\begin{split}
\ud \phi(x,t) &= (\ud t)  \frac{\delta H(\{\Pi_t,\phi_t\},a_t,b_t)}{\delta \Pi(x,t)} \, ,\\
\ud \Pi(x,t) &= - (\ud t)  \frac{\delta H(\{\Pi_t,\phi_t\},a_t,b_t)}{\delta \phi(x,t)}  
- (\ud t)\gamma \Pi(x,t) + \sqrt{\frac{2\gamma}{\beta \Delta x}} \ud B(x,t) \, ,
\end{split}
 \label{eqn:sde2}
\end{eqnarray}
where $\gamma$ is the parameter characterizing the inverse of the relaxation time and $\beta = (k_B T)^{-1}$. 
The standard Wiener process satisfies the following correlations:
$\uE\left[ \ud B(x_i,t) \right] = 0$ and 
$\uE \left[ (\ud B(x_i,t))(\ud B(x_j,t^\prime)) \right] = (\ud t) \delta_{i,j} \delta_{t,t^\prime}$, 
where $\uE[\,\,\,]$ represents the ensemble average for the Wiener process.
These fields are stochastic (not differentiable) 
and thus defined in the discretized space-time with the spatial interval $\Delta x$ and the time interval $\ud t$. 
For the sake of simplicity of notation, we use integral, functional derivative, etc in the following discussion, 
but these should be understood as the corresponding discretized operators,
\begin{eqnarray}
\int \ud x \longrightarrow (\Delta x) \sum \, , \,\,\,\, \,\,\,\,
\frac{\delta}{\delta \phi(x)} \longrightarrow \frac{1}{\Delta x} \frac{\partial}{\partial \phi(x)} \nonumber \, ,
\end{eqnarray} 
where the sum ($\sum$) runs over all discretized spatial points.
The continuum limit ($\Delta x , \ud t \rightarrow 0$) is performed at the last stage of calculations.
These stochastic differential equations are described in the rest frame of the heat bath and thus the equations are not manifestly Lorentz covariant.
See also the discussion in Ref.\ \cite{kkRBM}.

Let us define the functional phase space distribution of the scalar-field configuration by 
\begin{eqnarray}
f (\{\Pi, \phi\},t) 
=\int [\ud \Pi_0] \int [\ud \phi_0] \, f_0(\{\Pi_0, \phi_0 \}) \prod \uE \left[\delta (\Pi(x) - \Pi(x,t)) \delta (\phi(x) - \phi (x,t))\right] \, , \nonumber
\end{eqnarray}
where $\phi_0$ and $\Pi_0$ are the corresponding fields at an initial time $t_I$, satisfying the initial functional phase space distribution $f_0(\{\Pi_0, \phi_0 \})$.
To express the integral measures, we introduced the notation, $[\ud A] = \prod \ud A(x)$ for an arbitrary field $A(x)$,  
where the product ($\prod$) runs over all discretized spatial points.
Using Eq.\ (\ref{eqn:sde2}) and Ito's lemma \cite{book:gardiner}, the functional Fokker-Planck (Kramers) equation is thus given by
\begin{eqnarray}
\partial_t f (\{\Pi, \phi\},t)
&=& 
\int \ud x \, \left[ 
-\frac{\delta}{\delta \phi(x)} \left\{ \frac{\delta H(\{\Pi,\phi\},a_t,b_t)}{\delta \Pi(x,t)} f (\{\Pi, \phi\},t) \right\}
\right. \nonumber \\
&& \hspace{-1cm}\left.
+ \frac{\delta}{\delta \Pi(x)} \left\{ \left(
\frac{\delta H(\{\Pi,\phi\},a_t,b_t)}{\delta \phi(x,t)} 
+ \gamma \Pi (x,t) 
+ \frac{\gamma}{\beta} \frac{\delta}{\delta \Pi (x)} 
\right)   f (\{\Pi, \phi\},t) \right\} \right] \, . 
\end{eqnarray}
When the external parameters are constant, $a_t = a$ and $b_{t} = b$,
the stationary solution is given by the standard equilibrium distribution,
\begin{eqnarray}
f_{*} (\{ \Pi, \phi\},a,b) = \frac{1}{Z(a,b)}e^{-\beta H(\{\Pi,\phi\},a,b)} \, , \nonumber 
\end{eqnarray} 
with the partition function $Z(a,b) = \int [\ud \Pi] \int [\ud \phi] \, e^{-\beta H(\{\Pi,\phi\},a,b)}$.

The actual thermal relaxation process can be more complicated, 
but this model is expected to be a good approximation to describe the relaxation near equilibrium.
In fact, the non-equilibrium behavior predicted by the same type of model is experimentally confirmed \cite{blickle}.

In stochastic energetics, the heat absorbed by the scalar field is interpreted as the work done 
by the heat bath on the scalar field \cite{sekimoto}. 
The interactions between the scalar field and the heat bath are represented by 
the second and third terms on the right-hand side of the second equation in Eq.\ (\ref{eqn:sde2}), and these induce the change of the scalar field. 
The heat is thus defined by 
\begin{eqnarray}
\ud Q_t 
=
\int \ud x\, \left( - \gamma \Pi(x,t) + \sqrt{\frac{2\nu}{\Delta x}} \frac{\ud B (x,t)}{\ud t} \right)\circ \ud \phi (x,t) \, ,
\end{eqnarray}
where the Stratonovich definition of the product is 
$\ud B(x,t) \circ g(\phi(x,t)) = \ud B(x,t) \{ g(\phi(x,t)) + g(\phi(x,t+\ud t)) \}/2$ for an arbitrary smooth function $g(z)$ \cite{book:gardiner}.
It  is then straightforward to show that the expectation value of the heat is reexpressed as 
\begin{eqnarray}
\uE [\ud Q_t] = \ud {\cal E}(t) - (\uE[ \ud W_t] + \uE[\ud {\cal W}_t] ) \, , \label{1stlaw_d}
\end{eqnarray}
where the mean energy of this system is
\begin{eqnarray}
{\cal E}(t) 
= \int [\ud \Pi] \int [\ud \phi] \, f (\{\Pi, \phi\},t) H (\{\Pi, \phi\},a_t,b_t) 
\equiv 
\langle  H (\{\Pi, \phi\},a_t,b_t)  \rangle_t 
\, , \nonumber 
\end{eqnarray}
and the two works associated with the deterministic changes of $a_t$ and $b_t$ are, respectively, given by 
\begin{eqnarray}
\ud W_t =  \frac{\partial H(\{ \Pi_t,\phi_t \},a_t,b_t)}{\partial a_t} \ud a_t  \, ,  \,\,\,\,\,\,\, 
\ud {\cal W}_t = \frac{\partial H(\{ \Pi_t,\phi_t\},a_t,b_t)}{\partial b_{t}} \ud b_{t}  \, . \nonumber 
\end{eqnarray}
This represents the mean behavior of the energy balance in our fluctuating system 
and corresponds to the first law of thermodynamics. 
We have now two types of work. 
One ($\uE[ \ud W_t]$) indicates the change of the Hamiltonian induced by that of $a_t$ and hence 
is the work associated with the {\it spatial} volume change. 
The other ($\uE[\ud {\cal W}_t]$) is a new contribution representing the energy change induced 
by the control of $b_t$ and thus the work induced by the {\it functional} volume change.

To show the property analogous to the second law, we define the thermodynamical entropy by using Shannon's information entropy,
\begin{eqnarray}
S(t) = - k_B \int [\ud \Pi] \int [\ud \phi] \, f (\{\Pi,\phi\},t) \ln f (\{\Pi,\phi\},t) \, . \label{eqn:entropy}
\end{eqnarray}
Using this and the heat defined above, we find
\begin{eqnarray}
\lefteqn{\frac{\ud S (t)}{\ud t} - \frac{1}{T} \frac{1}{\ud t}\uE [\ud {Q}_t ]} && \nonumber \\
&=&
k_B 
\int [\ud \Pi] \int [\ud \phi]  \int \ud x  \, \frac{ \gamma \beta}{ f (\{\Pi, \phi\},t)}
\left[
\Pi (x,t) f (\{\Pi,\phi\},t) 
+
\frac{1}{\beta} \frac{\delta f (\{\Pi, \phi\},t)}{\delta \Pi(x)} 
\right]^2 \nonumber \\
&\ge& 0
\, .\label{eqn:2ndlaw}
\end{eqnarray}
The equality is satisfied when the functional phase space distribution is given by $f (\{\Pi, \phi\},t) = f_* (\{\Pi,\phi\},a_t,b_t)$, which 
is approximately satisfied in the quasi-static process where $\partial_t a_t$ and $\partial_t b_t$ are negligible.
Therefore, Eq.\ (\ref{eqn:2ndlaw}) corresponds to the second law of thermodynamics in our model.
It should be however noted that 
this is not equivalent to the second law because Eq.\ (\ref{eqn:entropy}) is defined even in non-equilibrium states.
The H-theorem in this system is shown using the Kullback-Lieber divergence (relative entropy) in Appendix \ref{app:1}.

Let us further investigate the equilibrium property. From the first and second laws, we can show 
\begin{eqnarray}
\int^{t_F}_{t_I} (\ud {\cal E}(t)  - T \ud S(t)) \ge  \int^{t_F}_{t_I}\uE[\ud W_t] + \int^{t_F}_{t_I}\uE [\ud {\cal W}_t] \, , \nonumber
\end{eqnarray}
where $t_F (> t_I)$ is a final time of the time evolution.
In the quasi-static limit where $f (\{ \Pi,\phi\},t) = f_* (\{ \Pi,\phi\}, a_t,b_t)$, we find that 
${\cal E}(t) = {\cal E}(a_t,b_t)$ and $S(t) = S(a_t,b_t)$, and thus 
the left-hand side is expressed using the thermodynamical potential $\Omega (T, a_t,b_t) = - \beta^{-1} \ln Z(a_t,b_t)$.
The above inequality is thus replaced by the equality:  
\begin{eqnarray}
\Omega(T,a_{t_F},b_{t_F}) - \Omega(T,a_{t_I},b_{t_I}) =  \int^{t_F}_{t_I}\uE[\ud W_t] + \int^{t_F}_{t_I}\uE [\ud {\cal W}_t] \, .
\end{eqnarray}
Because the left-hand side is determined only by the properties of the initial and final states and independent of the time dependence of $a_t$ and $b_t$, 
the quantities on the right-hand side should be given by 
\begin{eqnarray}
\uE[\ud W_t] =  \frac{\partial \Omega(T,a_{t},b_{t})}{\partial a_t} \ud a_t  \, ,\,\,\,\,\,\,\, 
 \uE[\ud {\cal W}_t] =  \frac{\partial \Omega(T,a_{t},b_{t})}{\partial b_t} \ud b_t \, . \nonumber 
\end{eqnarray}
For the sake of simplicity, let us consider $b_t =0$ in the thermodynamical limit.
The spatial volume is then represented by a function of $a_t$, $V = V(a_t)$, and 
thus the pressure $P$ is given by 
\begin{eqnarray}
P = - \frac{\partial \Omega(T,a_{t},0)}{\partial a_t} \frac{1}{(\ud V/\ud a_t)} 
= - \frac{\Omega(T,V(a_{t}),0)}{V(a_{t})} \, .
\end{eqnarray}
In the second equality, we used that the thermodynamical potential is an extensive quantity in the thermodynamical limit. 
This expression of the pressure is consistent with the well-known result in the finite-temperature field theory \cite{FTFT}.

To consider thermodynamics associated with the functional volume change, we set $a_t=0$ in the thermodynamical limit. 
Applying a similar discussion, 
we can define the pressure associated with the change of the functional volume $V_F (b_t)$ by 
\begin{eqnarray}
P_F 
= - \frac{\partial \Omega(T,0,b_{t})}{\partial b_t} \frac{1}{(\ud V_F/\ud b_t)}  
= - \frac{\Omega(T,0,V_F (b_{t}))}{V_F(b_{t})} \, . 
\label{eqn:fpressure}
\end{eqnarray}
Here we used again that the thermodynamical potential 
is a linear function of the extensive quantity, $V_F(b_t)$.
Therefore the first law in the quasi-static infinitesimal process is 
\begin{eqnarray}
\ud {\cal E} - T \ud S =  - P_F \ud V_F  \, , \label{eqn:tdr1}
\end{eqnarray}
and the new thermodynamical quantities $P_F$ and $V_F$ are introduced consistently with the first (\ref{1stlaw_d}) and second laws (\ref{eqn:2ndlaw}).

\begin{figure}[h]
\begin{center}
\includegraphics[scale=0.4]{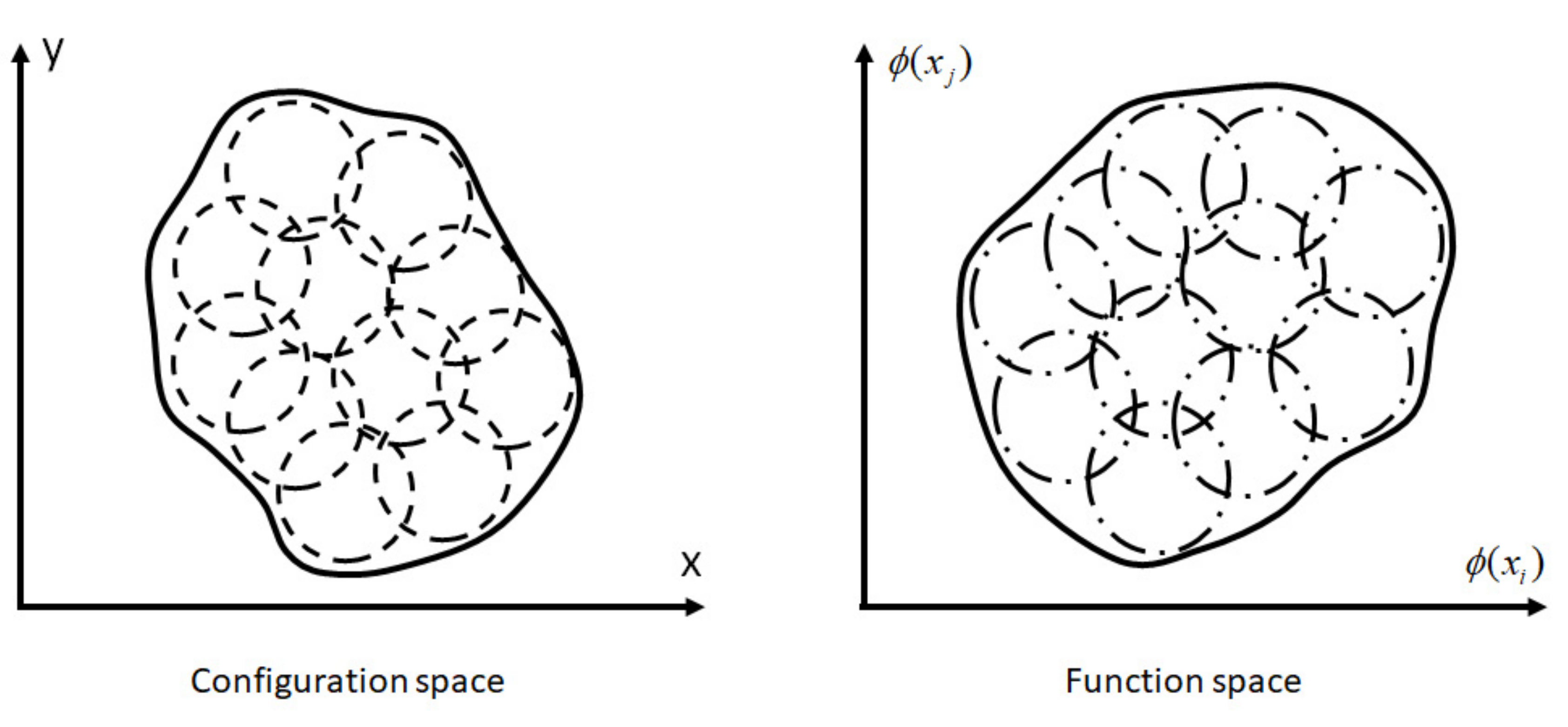}
\caption{The schematic figures of the distributions of the fluids in the spatial configuration space (left) and the function space (right). 
The left panel shows the decomposition of the fluid by fluid elements (the dashed circles) and 
the right panel represents the decomposition by functional fluid elements (the dot-dashed circles). 
The position $x_i$ is different from the position $x_j$ on the right panel.}
\label{fig:f_elements}
\end{center}
\end{figure}

As an application of the derived results, we consider the generalization of the local equilibrium ansatz to the function space.
Let us consider a fluid composed of the 1+1 dimensional real scalar field which distributes in the function space with a certain probability. 
As the fluid composed of particles is represented by the ensemble of fluid elements, 
the fluid in the function space is decomposed into the functional fluid elements as shown in Fig.\ \ref{fig:f_elements}.
The local equilibrium ansatz in the function space means that the internal states of these functional fluid elements 
are approximately given by thermal equilibrium states.
We show that there exists a simple ideal fluid model in the function space, which is consistent with the derived thermodynamical relation (\ref{eqn:tdr1}).

As the functional hydrodynamical variables, 
the probability distribution of the scalar-field configuration $\rho_F [\phi,t]$, the fluid velocity functional $v [x, \varphi, t]$ and 
the internal energy distribution $\varepsilon_F [ \varphi ,t ]$ are chosen.
Functional ideal hydrodynamics is obtained by applying the variational principle to the following Lagrangian \cite{fhydro}, 
\begin{eqnarray}
L 
&=&  \int [\ud \phi_0] \, 
\rho_0 [\phi_0] \left[   \int \ud x\,  \left\{ \frac{1}{2} (\partial_t \phi (x,t))^2 - \frac{1}{2} (\partial_x \phi (x,t))^2 \right\} 
-  \frac{\varepsilon_{F}  [ \phi ,t ]}{\rho_F [\phi,t]} \right]\, ,\label{eqn:lag}
\end{eqnarray}
where $\rho_0 [\phi_0] $ denotes the initial probability distribution of the scalar-field configuration.
To be more accurate, the fluid describes only the macroscopic motions and thus should be represented 
by an effective scalar field $\phi_{eff}(x,t)$ which is obtained by the coarse-graining associated with rapid spatial and temporal oscillations of $\phi(x,t)$. 
This coarse-graining is however not explicitly considered below because the derivation of the equations is out of the scope of the purpose of this  letter. 
See Ref.\ \cite{fhydro} for details.

The equations of functional ideal hydrodynamics are given by 
\begin{eqnarray}
\begin{split}
\frac{\ud}{\ud t_\phi} \rho_F [ \phi, t ] 
&= - \rho_F [ \phi, t ] \int \ud x\,  \frac{\delta v [ x, \phi,t ] }{\delta \phi(x)}    \, ,
\\
\frac{\ud}{\ud t_\phi} v [x, \phi, t]
&=  \partial^2_x \phi (x) 
-  \frac{1}{\rho_F [\phi,t]} \frac{\delta P_F (\varepsilon_F, \rho_F)}{\delta \phi(x)}
\, , \\
%
\frac{\ud}{\ud t_\phi} \varepsilon_F [ \phi ,t ] 
&= 
-  \{ \varepsilon_F [\phi,t]  + P_F (\varepsilon_F, \rho_F) \}  \int \ud x\,  \frac{\delta  v [x, \phi,t] }{\delta \phi(x)} 
\, ,
\end{split}
\label{eqn:f-hydro}
\end{eqnarray}
where the material derivative in the function space is defined by 
\begin{eqnarray}
\frac{\ud}{\ud t_\phi} = \partial_t  + \int \ud x\, v [x, \phi, t]\frac{\delta}{\delta \phi(x)} \, . \nonumber
\end{eqnarray}
The first equation is the conservation of the probability. 
The second equation is related to the momentum conservation and obtained by the variation of the above Lagrangian.
The third equation is obtained so that the fluid energy, which is given by the Hamiltonian derived from Eq.\ (\ref{eqn:lag}), is conserved.
The structures of these equations can be regarded as the functional generalizations of the non-relativistic ideal-fluid equations.

Assuming that the volume of one functional fluid element $V_F$ is characterized by $\rho^{-1}_F$,
Eq.\ (\ref{eqn:tdr1}) is reexpressed in terms of the functional hydrodynamical variables as 
\begin{eqnarray}
\ud \left( \frac{\varepsilon_F}{\rho_F} \right) - T \ud \left(\frac{s_F}{\rho_F}\right) = -P_F \ud\left( \frac{1}{ \rho_F} \right) \, , \label{eqn:tdr2}
\end{eqnarray}
where $\varepsilon_F = {\cal E}/V_F$, and $s_F = S/V_F$ is the entropy distribution in the function space.
The functional local equilibrium ansatz requires that this relation is satisfied to the internal state of each functional fluid element.
It should be emphasized that this thermodynamical relation is applied to the rest frame of the functional fluid element.
When a fluid element at $x$ of standard hydrodynamics moves with a velocity $v(x,t)$, thermodynamics is applied to the internal state observed from the co-moving frame of $v(x,t)$ .
In a similar fashion, when a functional fluid element at $\phi(x)$ in the function space moves with the velocity functional $v [x, \phi, t]$,  
Eq.\ (\ref{eqn:tdr2}) is applied to the internal state observed from the co-moving frame of $v [x, \phi, t]$.

If the above ansatz is appropriate, the entropy of the fluid is a conserved quantity because Eq.\ (\ref{eqn:f-hydro}) 
describes the ideal fluid.
The equation of $s_F$ is calculated from Eq.\ (\ref{eqn:tdr2}) by replacing $(\ud)$ with the functional material derivative $(\ud/\ud t_\phi)$ 
which corresponds to the differential in the co-moving frame of the functional fluid element, 
\begin{eqnarray}
\partial_t s_F [ \phi, t ]  = - \int \ud x\, \frac{\delta}{\delta \phi(x)} \left\{ s_F [ \phi, t ]  v [ x, \phi,t ] \right\} \, .
\end{eqnarray}
This is the equation of continuity in the function space 
and thus it is easy to see that the total entropy of the fluid $S= \int [\ud \phi]  s_F [ \phi, t ]$ is conserved. 
This consistency suggests the applicability of thermodynamics to the abstract manipulation in the function space.

There are several remarks.
In the stochastic model defined by Eq.\ (\ref{eqn:sde2}), we consider the infinitely large heat bath where 
the scalar field $\phi(x,t)$ interacts with the heat bath anywhere with the same relaxation coefficient $\gamma$.
Thus the expectation value of the heat is infinite in general,
\begin{eqnarray}
\frac{1}{\ud t}\uE [\ud Q_t] = \uE \left[ \int \ud x\, \left( -\gamma \Pi^2(x,t) + \frac{\gamma}{\beta \Delta x}\right)  \right] \, . \nonumber
\end{eqnarray}
This divergence, however, should not be simply ignored because it plays an important role to obtain the second law (\ref{eqn:2ndlaw}).

The functional pressure $P_F$ should be calculated from the finite-temperature field theory using the definition (\ref{eqn:fpressure}).
For this, we have to define the functional volume $V_F$ but there is no established definition. 
This might be characterized by a field correlation function such as 
$\sqrt{ \langle \prod (\phi(x) - \langle \phi (x) \rangle_t)^2 \rangle_t }$.
The functional volume change is related to, for example, the physics of the mass shift of hadrons 
induced by medium effects. 
The detailed studies of $V_F$ and $P_F$ may thus help us to improve our understanding to 
the medium effects in subatomic physics \cite{tsushima,gubler}.
The work related to $P_F$ and $V_F$ is considered also in Ref.\ \cite{deffner} to discuss the Jarzynski relation in quantum field theory.

In the present approach, one functional fluid element is defined by using the information of the field in the whole space at each instant.
Such a definition of the functional fluid element may conflict with relativistic causality.
The model considered here is however a simplest version and it is possible to consider the generalization. 
For example, the energy distribution depends not only on the field configuration but also on the spatial position ${\varepsilon (x,\{\Pi,\phi\},t)}$. 
Moreover it is possible to introduce the effects of the quantum-field theoretical fluctuation and viscosity to functional hydrodynamics 
by applying the stochastic variational method \cite{fhydro,svm,svm-field}.
These are left as future tasks.

\vspace*{1cm}
T. Koide thanks to K. Tsushima for useful comments for the references.  
The authors acknowledge the financial supports by CNPq (No.\ 305654/2021-7, \, 303246/2019-7),
FAPERJ and CAPES. 
A part of this work has been done under the project INCT-Nuclear Physics and Applications (No.\ 464898/2014-5).

\appendix

\section{H-theorem} \label{app:1}

When $a_t=a$ and $b_t=b$, we define the Kullback-Leibler divergence by
\begin{eqnarray}
J(f|f^*) =  \int [\ud \phi]\,  f(\{\Pi,\phi\},t) \ln \frac{f(\{\Pi,\phi\},t)}{f_*(\{\Pi,\phi\},a,b)} \, .
\end{eqnarray}
This satisfies the H-theorem, 
\begin{eqnarray}
\frac{\ud J(f|f^*)}{\ud t }
= 
-\frac{\gamma}{\beta} \int [\ud \phi]  \, f (\{\Pi,\phi\},t) \int \ud x\,
 \left( \frac{\delta }{\delta \Pi(x)} \ln \frac{f(\{\Pi,\phi\},t)}{f_*(\{\Pi,\phi\},a,b)}  \right)^2 \le 0 \, .
\end{eqnarray}
In the Boltzmann equation, the quantity which satisfies the H-theorem is interpreted as the entropy 
but this interpretation is not applicable to Brownian motion.

\end{document}